\documentclass[prl,amsmath,amssymb,superscriptaddress,twocolumn]{revtex4}
\usepackage{graphicx}
\usepackage {amsmath}
\usepackage {amsfonts}
\usepackage {amssymb}
\usepackage {mathrsfs}
\usepackage {bm}
\usepackage {mathbbol}
\usepackage[usenames]{color}

\newcommand{\br}{{\bf r}}
\newcommand{\bx}{{\bf x}}

\newcommand{\bk}{{\bf k}}

\newcommand{\bK}{{\bf K}}

\newcommand{\bR}{{\bf R}}

\newcommand{\ms}{{m^{\ast}}}

\newcommand{\eps}{\epsilon}

\DeclareMathAlphabet{\mathpzc}{OT1}{pzc}{m}{it}

\newcommand {\be}{\begin{eqnarray}}
\newcommand {\ee}{\end{eqnarray}}
\newcommand {\bw}{\begin{widetext}}
\newcommand {\ew}{\end{widetext}}

\newcommand {\rmd} {{\rm d}}

\newcommand{\Tr}{{\rm Tr}}
\newcommand{\bbone}{{\mathbb 1}}

\newcommand{\omegan}{{\omega_n}}

\begin{document}
\title{Topology and symmetry breaking in ABC trilayer graphene}
\author{Vladimir Cvetkovic}
\author{Oskar Vafek}
\affiliation{National High Magnetic Field Laboratory and Department
of Physics,\\ Florida State University, Tallahassee, FL 32306, USA}

\date{\today}

\begin{abstract}
The effects of topology and electron-electron interactions on the phase diagram of
ABC stacked trilayer graphene (TLG) at the neutrality point are investigated
within a weak coupling renormalization group approach. We find that the leading instability of TLG with only a
forward scattering density-density
interaction is towards a mirror-breaking gapless state. Addition of a small, but finite back scattering favors gapped phases,
allowing us to make connections to the existing experiments on TLG. We identify a fundamental symmetry difference
between TLG and bilayer graphene (BLG) which is responsible for disfavoring nematic states in TLG under
the same conditions that favor nematic states in BLG.
\end{abstract}

\maketitle

Interest in the theory of metal-insulator transitions in divalent materials \cite{HalperinRice68} has undergone a
resurgence as of late in no small part due to the discovery of graphene \cite{GeimNovoselov2004,GeimNovoselov2005,Kim2005}.
There is growing experimental evidence for electron-electron (e-e) interaction-induced excitonic instabilities in AB stacked
bilayer (BLG) \cite{YacobyScience2010,YacobyPRL2010,Mayorov2011,VelascoNatNano2012,VeliguraPRB2012,FreitagPRL2012,BaoArXiv}
and the ABC stacked trilayer graphene (TLG) \cite{LauTLGNatPhys2011}. 
Theoretically, based on symmetry alone \cite{HalperinRice68}, TLG is qualitatively different from single layer graphene (SLG) and BLG.
The valence band (VB) and the conduction band (CB) in SLG and BLG are
members of the {\em same} irreducible representation (IR) of the symmetry group
at the $\pm {\bf K}=\pm\frac{4\pi}{3\sqrt{3}a}\hat{\bx}$ points in the Brillouin zone.
On the other hand, the VB and the CB of TLG belong to {\em different} IR's such that
the interband matrix element of the momentum operator vanishes at ${\pm \bK}$.

SLG, BLG, and TLG are further distinguished in the topological character of their VB and the CB Bloch
states as they wind along a closed {\bf k}-space loop around the $\pm \bK$ points. Their winding numbers, which are
$1$, $2$ and $3$ respectively \cite{McCannFalkoPRL2006,KoshinoMcCannPRB2009}, can be related to the experimentally
measured steps in the quantized Hall conductivity as the carrier concentration
is tuned from hole- to electron-like \cite{GeimNovoselov2005,BLGwindingExperiment,ZaliznyakTLG2011}. This
non-trivial topology guarantees at least three
Dirac Fermions (DF) near, but not exactly at, each of $\pm\bK$ of BLG and TLG, placed along the 3-fold symmetry lines in the BZ. Since one can
adjust the microscopic parameters of the model {\em without lowering the symmetry} so as to
either rotate the three DF's or to introduce an additional six DF's near each $\pm\bK$, the
Dirac degeneracies in TLG cannot be established based on symmetry alone \cite{noDFs}.
This raises the possibility of zero temperature phase transitions between different, but symmetry indistinguishable, thermodynamic phases.

\begin{figure}[t]
\begin{center}
\includegraphics[width=0.5\textwidth]{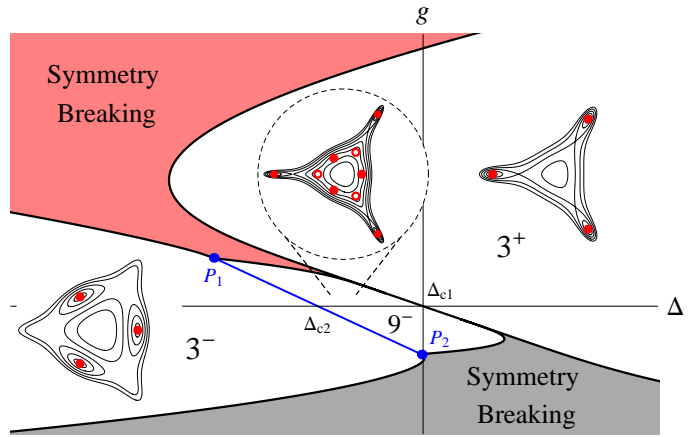}
\end{center}
  \caption{Schematic zero temperature phase diagram for ABC stacked trilayer graphene at the neutrality point. $g$ is the strength of the
    electron-electron interactions and $\Delta$ the conduction and valence band mixing at the symmetry points $\pm \bK$
    (see Eq.\ (\ref{eq:H kinetic})). In the non-interacting case $\Delta_{c1,c2}$ mark the quantum phase transitions between
    distinct, but symmetry equivalent, states $3^+$, $9^-$, and $3^-$. Their schematic contours of constant energy near the $\bK$ point with
    Dirac points (red circles) are shown. The (blue) phase boundary between $3^-$ and $9^-$ terminates at two tricritical points $P_{1}$
    and $P_2$. At any finite $g$ the direct transition between $3^+$ and $9^-$ is avoided and replaced by transitions into a symmetry broken
    phase, whose width becomes exponentially small as $g\rightarrow 0$.}\label{fig:schematic phase diagram}
\end{figure}

In this work we systematically investigate the interplay of such topological effects and the e-e interactions in TLG.
We employ a powerful theoretical machinery based on symmetry and renormalization group (RG), which leads us to the basic
structure of the general phase diagram shown in Fig.\ \ref{fig:schematic phase diagram} together with the list of, and the conditions for, the most
dominant symmetry breaking (SB) phases. As a biproduct, our symmetry analysis allows us to readily explain why 3-fold rotational
SB gapless states in TLG considered recently within Hartree-Fock approximation are indeed
unfavorable \cite{JungMacDonald2012}, unlike in BLG \cite {VafekYangPRB2010,LemonikPRB2010,LemonikPRB2012,
CvetkovicThrockmortonVafek2012}, while noting that there are 3-fold rotationally symmetric
gapless states, not considered in Ref.\ \onlinecite{JungMacDonald2012}, which break the mirror symmetry of the lattice and
are stable in the limit of pure forward scattering.
Nevertheless, small, {\it but finite} backscattering (i.e., intervalley scattering) makes gapless states unstable and promotes gapped phases.
Our theory focuses from the outset on the most important low energy degrees of freedom 
and extends significantly beyond the recent functional RG treatment \cite{MScherer2012} that
did not take into account RG relevant symmetry allowed terms and treated only up to nearest neighbor lattice interactions.

\begin{figure}[t]
  \includegraphics[width=0.49\textwidth]{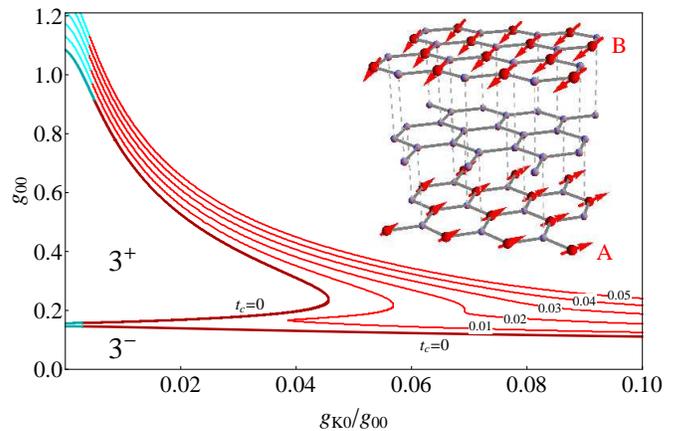}
  \caption{The influence of back scattering on the transition temperature and symmetry breaking channel. The color of isotherms corresponds to the leading
    instability: mirror-breaking (cyan), anti-ferromagnet (red). In $3^\pm$ regions no symmetry breaking occurs,
    even at $T=0$. Inset: a TLG lattice with the AF order. }\label{fig:isotherms}
\end{figure}

Because the experimentally observed \cite {WBaoNatPhys} energy (few $meV$) scales associated with any ordering are small compared
to the relevant band splitting scales (few $100$'s $meV$), the theory of the TLG needs to take into account only
the modes which lie close to the Fermi level. 
The effective low energy Hamiltonian written in the vicinity of the $\pm\bK$ points is
\begin{eqnarray}
  H=H_0+H_{int}. \label {Hamiltonian}
\end{eqnarray}
Setting $\hbar=k_B=1$, the symmetry allowed terms in the non-interacting Hamiltonian are
\begin{eqnarray}\label{eq:H kinetic}
  H_0=\sum_{\bk,\sigma=\uparrow,\downarrow}\psi_{\bk\sigma}^{\dagger}
    \left(h^{A_{1g}^+}_{\bk}+h^{A_{1u}^-}_{\bk}+h^{A_{2u}^-}_{\bk}\right)\psi_{\bk\sigma}, \label {H0}
\end{eqnarray}
where
$h^{A_{1g}^+}_{\bk}=(-\eps_F+\frac{1}{2\delta m^*}\bk^2)\bbone_4+(\Delta+\frac{1}{2m^*}\bk^2)\bbone_2\sigma_1$,
$h^{A_{1u}^-}_{\bk}=R_1k_x(k_x^2-3k_y^2)(R_0\tau_3\bbone_2+\tau_3\sigma_1)$ and
$h^{A_{2u}^-}_{\bk}=R_2k_y(3k^2_x-k^2_y)\bbone_2\sigma_2$. The Pauli matrices $\tau$ and $\sigma$ act in the valley and layer space respectively.
For each spin projection, the Fermi annihilation operators have been put into a four component vector $\psi_{\bk\sigma}=\left(c^A_{\bk+\bK\sigma},c^B_{\bk+\bK\sigma},c^A_{\bk-\bK\sigma},c^B_{\bk-\bK\sigma}\right)^T$
and $|\bk|<\Lambda$ where $\Lambda\ll|\bK|$. $A$ and $B$ refer to two undimerized sites on layers 1 and 3 (see Fig.\ \ref {fig:isotherms} inset).
There is some uncertainty in the actual values and {\em signs} of the coefficients $\ms$ and
$\Delta$ \cite {SlonczewskiMcClure, PartoensPRB2006,FZhangPRB2010}.
In this Letter we choose $\ms > 0$ and sweep over a range of both positive and negative $\Delta$'s. A situation
where $\ms<0$ can be mapped onto the $\ms > 0$ problem by simultaneous change in sign of $\ms$ and $\Delta$. Therefore,
the results presented here cover every possible arrangement of parameter signs.

For $\Delta>\Delta_{c1}=0$ there are three anisotropic DF's along three symmetry lines, and we label
this state $3^+$ (see Fig.\ \ref{fig:schematic phase diagram}). For
$\Delta<\Delta_{c2}<\Delta_{c1}$ there are also three DF's but rotated by $180^\circ$ relative to $3^+$. We label this state $3^-$.
For $\Delta_{c2}<\Delta<\Delta_{c1}$, and in the presence of {\em particle hole symmetry}, which we assume to hold from now on,
$R_0=0$, and $1/\delta m^*=0$, and the spectrum contains nine DFs; we label this state
$9^-$ (see Fig.\ \ref{fig:schematic phase diagram}) \cite{ph}. Because at
$T=0$ it is impossible to transition between any two of $3^+$, $3^-$, and $9^-$ states without
encountering a non-analyticity in the ground state energy, even in this non-interacting case, there are
two $T=0$ quantum phase transitions.
The transition at $\Delta_{c1}$ is $2^{nd}$ order and the one at $\Delta_{c2} = -1/(54 (R_1)^2 \ms^3 )$ is $3^{rd}$ order \cite{ehrenfest}.

We find that, for arbitrarily small e-e interactions, the $2^{nd}$ order phase
transition, initially at $\Delta_{c1}$, is avoided and replaced by two continuous phase transitions into a spontaneously broken
symmetry phase (see Fig.\ \ref{fig:schematic phase diagram}). The nature of the SB phase depends on the type and the strength
of the e-e interactions and we discuss it in detail later in the text. On the other hand, the $3^{rd}$ order
phase boundary point at $\Delta_{c2}$ turns into a critical line $\Delta_{c2}(g)$ for a finite range
of weak interactions $g$, before it is terminated in quantum tricritical points. 

We construct $H_{int}$ using
the irreducible representations (IR's) of the space symmetry group $P\bar 3 m 1$ \cite {LatilPRL2006}.
The most relevant scattering processes are described by the quartic contact terms. These are the products of two 
bilinears $g_{MN} \left ( \psi_{\alpha}^\dagger(\br) M_{\alpha\beta} \psi_{\beta}(\br) \right )
\left ( \psi_{\lambda}^\dagger(\br) N_{\lambda\rho} \psi_{\rho}(\br) \right )$, where, neglecting the
small spin-orbit coupling, the spin $SU(2)$ symmetry is assumed to be present, and so in the singlet channel
$M_{\alpha\beta} = \tau_\mu \sigma_\nu \delta_{\alpha \beta}$ while in the triplet channel
$M_{\alpha\beta} = \tau_\mu \sigma_\nu \vec \sigma_{\alpha \beta}$. Similar expressions hold for
$N_{\lambda\rho}$ and $\mu,\nu=0,1,2,3$; ($\tau_0$ or $\sigma_0$ $=\bbone_2$). When the quartic terms are
integrated over $\br$, the symmetries of the lattice require that the product of two IR's to which $M$ and $N$ belong is the
trivial IR. Further simplification comes from the fact that all the spin triplet-triplet terms can be written in terms of singlet-singlet
terms using Fierz identities \cite{VafekPRB2010,LemonikPRB2012}. Therefore, the most general contact quartic interaction
terms preserving the time reversal symmetry, lattice symmetry, and spin $SU(2)$
symmetry are
\be
&& H_{int} = \frac{4\pi}{| \ms |} \int \rmd \br \bigg \lbrack \sum_{\mu \nu} g_{\mu \nu} 
    \Big ( \sum_{\sigma} \psi_\sigma^\dagger(\br) \tau_\mu \sigma_\nu \psi_\sigma(\br) \Big )^2 \label{eq:Hint} \\
    && +2 \sum_{\mu} \tilde g_\mu \sum_{\sigma, \sigma'}
    \Big ( \psi_\sigma^\dagger(\br) \tau_\mu \bbone_2 \psi_\sigma(\br) \Big ) \Big ( \psi_{\sigma'}^\dagger(\br) \tau_\mu \sigma_1 \psi_{\sigma'}(\br) \Big )
    \bigg \rbrack, \nonumber
\ee
where $g_{1 \nu} = g_{2 \nu} \equiv g_{\bK \nu}$, for each $\nu$, and $\tilde g_1 = \tilde g_2 \equiv \tilde g_\bK$, and we
inserted the factor of $8\pi/ | \ms |$, which makes $g$'s dimensionless, for
convenience.
The total number of independent, symmetry allowed, couplings is $15$. We discuss the specific model choices of $g$'s further in the text.

To find the phase diagram, we evaluate the partition function $Z=\mbox{Tr}\left[e^{-H/T}\right]$ using coherent
state path integrals and the Wilson renormalization group (RG) procedure \cite{ShankarRMP1994}
which, for BLG, has been spelled out in detail in Ref.\ \onlinecite{CvetkovicThrockmortonVafek2012}.
As the modes within the thin momentum shell $1-\rmd \ell < |\bk|/\Lambda < 1$ and {\em any} Matsubara frequency
$\omega_n=(2n+1)\pi T$ are integrated out in powers of small $g$'s, the remaining modes, $\psi_{\bk\sigma}^<(\omega_n)$
and their momenta are rescaled such that the new $|\bk|$ cutoff is again $\Lambda$ and we choose the coefficient
of $\frac{\bk^2}{2m^*}\sigma_1\bbone_2$ in $h^{A_{1g}^+}_{\bk}$ to remain constant. To $1$-loop order we obtain the following RG flow equations
\begin{eqnarray}
  \frac {\rmd \ln t}{\rmd \ell} &=& -2 \frac {\rmd \ln r}{\rmd \ell} = 2, \label{eq:trflow} \\
  \frac {\rmd \delta}{\rmd \ell} &=&\!2 \delta\!+\!
    \!\sum_{\mu}\!\bigg \lbrack \tilde{g}_{\mu} \tilde{C}_\mu\!+ \!\sum_{\nu} g_{\mu \nu} C_{\mu\nu}
    \Phi_0\! \left ( t_{\ell}, \delta_\ell, r_{\ell} \right)\!\bigg\rbrack,  \label{eq:deltaflow} \\
  \frac {\rmd \tilde{g}_{\mu}}{\rmd\ell} &=& \sum_{\alpha \beta} \sum_{\kappa \lambda \kappa' } \tilde A_{\mu (\alpha \beta)}^{\kappa \lambda, \kappa'}
    g_{\kappa \lambda} \tilde g_{\kappa'} \Phi^{(\alpha \beta)} \! \left ( t_{\ell}, \delta_{\ell}, r_{\ell} \right), \label{eq:gtildeflow}
\ee
\be
  \frac {\rmd g_{\mu\nu}}{\rmd \ell} &=& \sum_{\alpha \beta} \bigg \lbrack \sum_{\kappa \lambda \kappa' \lambda'}
    A_{\mu\nu (\alpha \beta)}^{\kappa \lambda, \kappa' \lambda'} g_{\kappa \lambda} g_{\kappa' \lambda'} + \nonumber \\
  &&  \qquad \sum_{\kappa \kappa'} \bar A_{\mu\nu (\alpha \beta)}^{\kappa \kappa'} \tilde g_\kappa \tilde g_{\kappa'}  \bigg \rbrack
    \Phi^{(\alpha \beta)} \! \left ( t_{\ell}, \delta_{\ell}, r_{\ell} \right),\label{eq:gflow}
\end{eqnarray}
where $t_{\ell=0}=T/\frac{\Lambda^2}{2\ms}$, $\delta_{\ell=0}=\Delta/\frac{\Lambda^2}{2\ms}$ and for simplicity
we set $R_1=R_2=R$ in Eq.\ (\ref{eq:H kinetic}) giving $r_{\ell=0}=\left(2\ms\Lambda\right) R$.
The expressions for the coefficients $C_{\mu\nu}$, $\tilde{C}_{\mu}$, $A$, $\bar A$, and $\tilde A$'s are
too long to present here, but they are given in the Supplementary Material (SM). $\Phi$ functions are
\be
  \Phi_0 \left ( t, \delta, r \right)\!\! &=&\!\! \frac T 2 \sum_\omegan \int_0^{2\pi} \frac {\rmd \varphi}{2 \pi}
    \Tr \left \lbrack G_{\Lambda, \varphi} (\omegan) \bbone \sigma_1 \right \rbrack, \label {Phi0} \\
  \Phi^{(\mu \nu)}  \left ( t, \delta, r \right)\!\!  &=&\!\! \frac T {16} \sum_\omegan \int_0^{2\pi} \frac {\rmd \varphi}{2 \pi}
     \left(\Tr \left\lbrack G_{\Lambda, \varphi} (\omegan) \tau_\mu \sigma_\nu \right \rbrack\right)^2 . \label {Phiab}
\ee
In the above, the one particle Greens function, given by
$G^{-1}_{\bk} (\omegan)=-i\omega_n+h^{A^+_{1g}}_{\bk}+h^{A^-_{1u}}_{\bk}+h^{A^-_{2u}}_{\bk}$, is evaluated at
the circle $|\bk|=\Lambda$ parameterized by $\varphi$.
\begin{figure}[t]
\begin{tabular}{cc}
\includegraphics[width=0.21\textwidth]{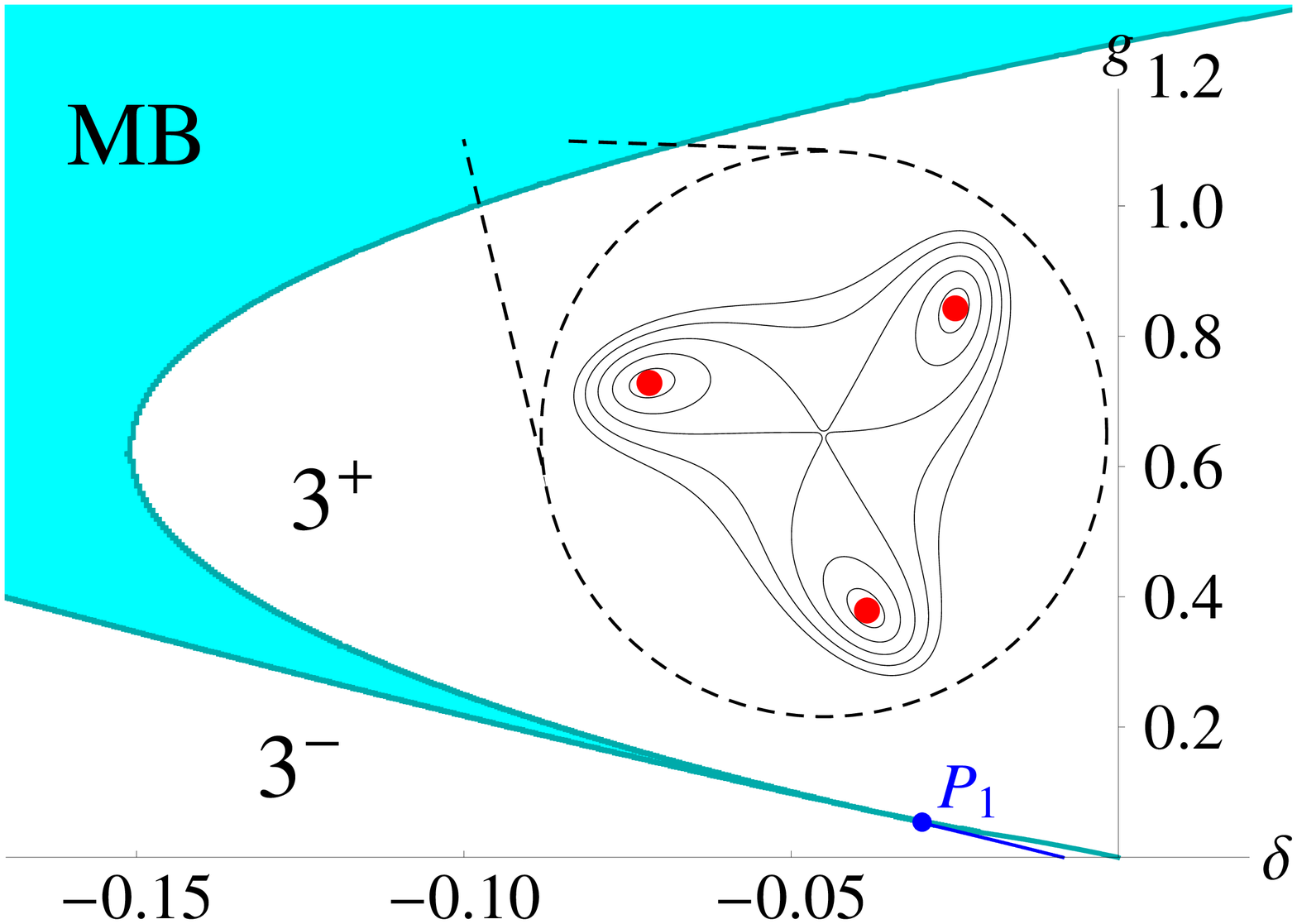} (a)&
\includegraphics[width=0.21\textwidth]{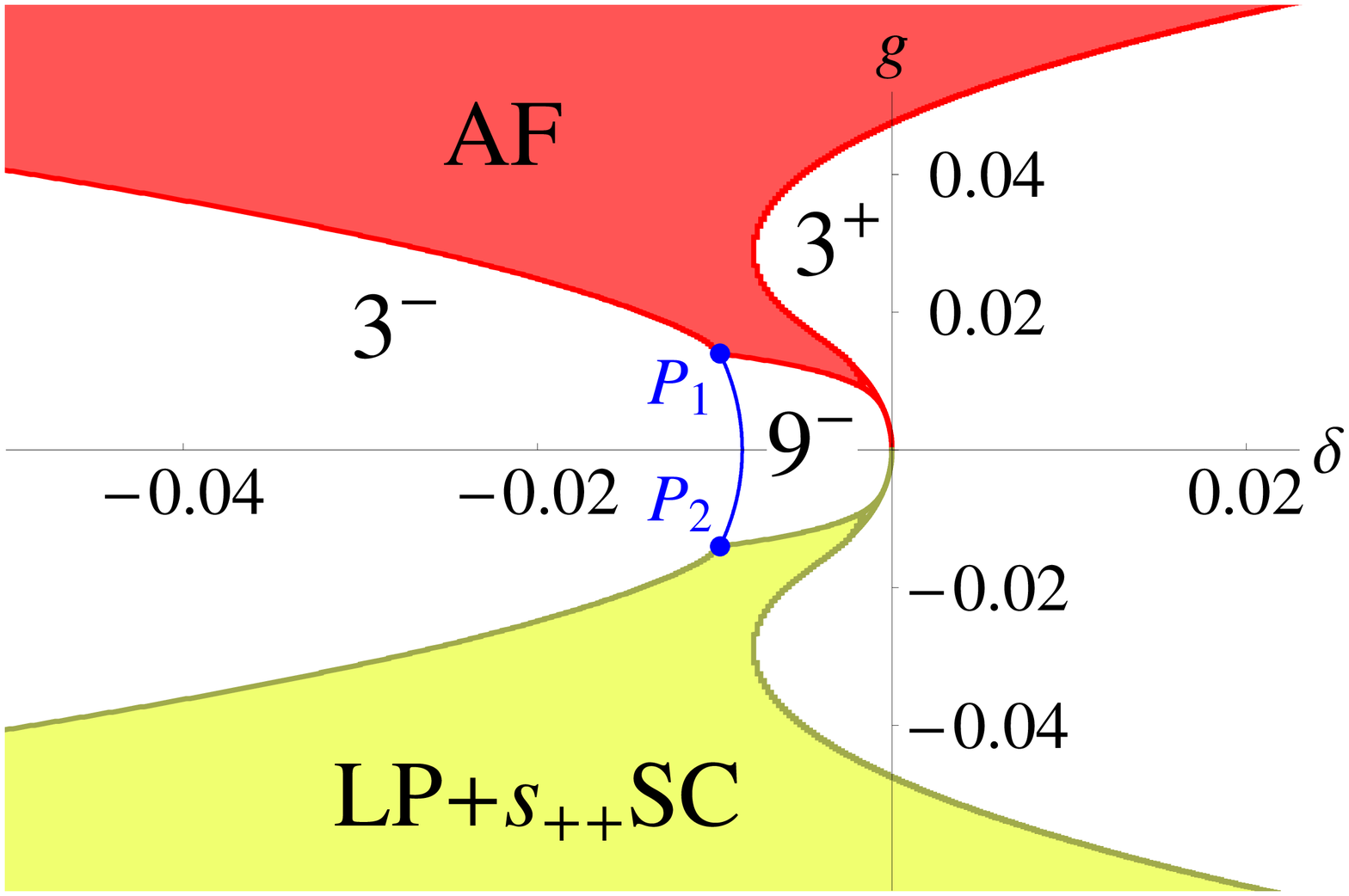} (b)\\
\includegraphics[width=0.21\textwidth]{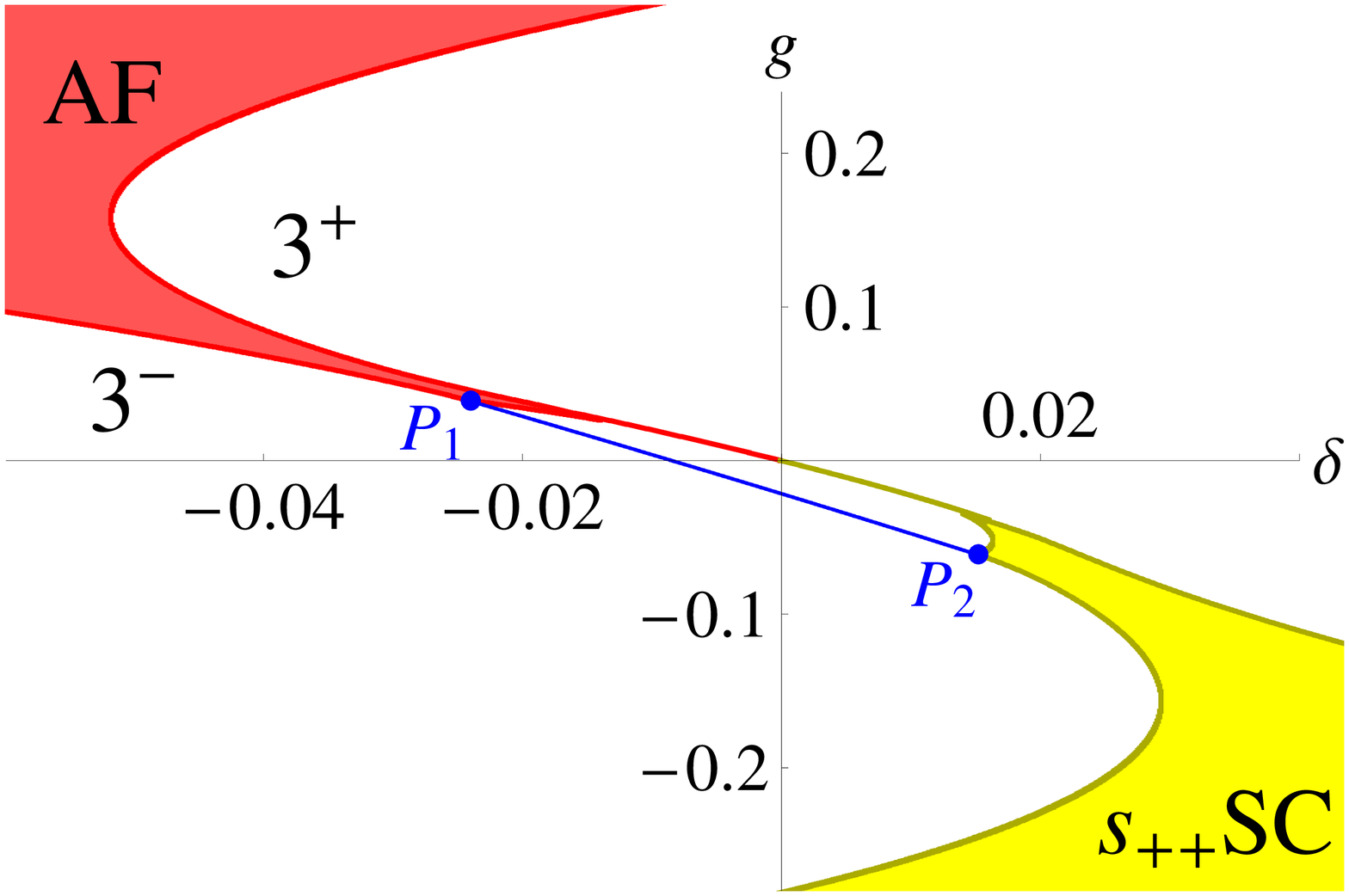} (c)&
\includegraphics[width=0.21\textwidth]{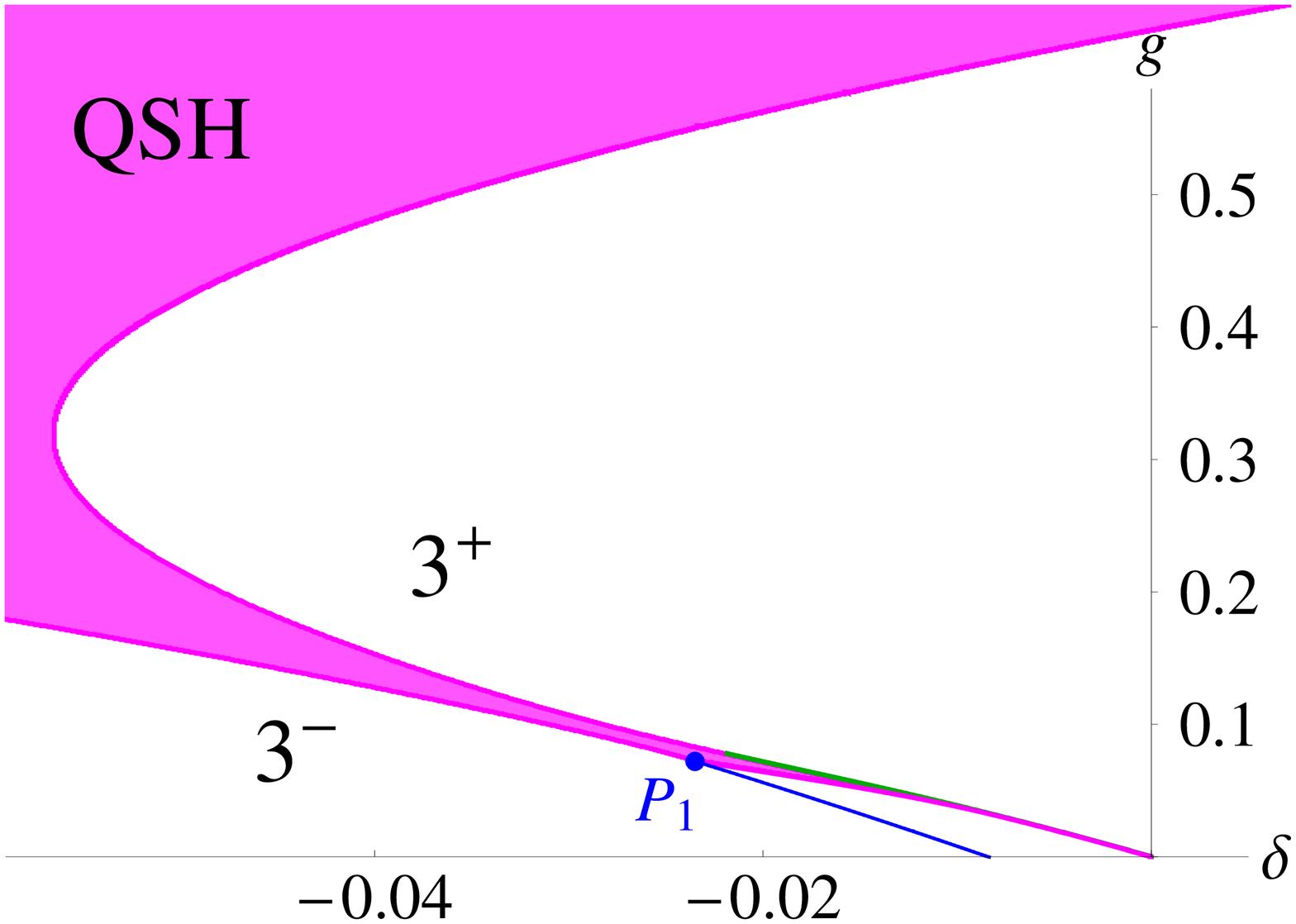} (d)
\end{tabular}
\caption{$T=0$ phase diagrams for the three interaction models for TLG studied here.
(a) $g_{00}$-only, (b) weak coupling Hubbard model, (c+d) Screened Coulomb interaction.}\label{fig:phase diagrams}
\end{figure}

While the details of the flow described by the above equations must be solved numerically, the general trends can be readily understood.
From Eq.\ (\ref{eq:trflow}), $t_{\ell}=t_{\ell=0}e^{2\ell}$ and $r_{\ell}=r_{\ell=0}e^{-\ell}$.
Starting from some fixed initial values of $g$'s and sufficiently high temperature $t_{\ell=0}$, the exponential decay of
the $\Phi$'s with growing $\ell$
quickly halts the growth of the magnitude of the $g$'s, which saturate to finite values as $\ell\rightarrow \infty$.
Such a flow corresponds to a phase with no spontaneous SB.
Holding the values of the initial $g$'s fixed and reducing $t_{\ell=0}$, there is an increase in the values of $\ell$
where $\Phi$'s halt the flow, and the limiting magnitude of the $g$'s increases as $\ell\rightarrow\infty$. Whether the $g$'s
eventually diverge upon further lowering of $t_{\ell=0}$ depends on the initial values of the remaining parameters. The
highest $t_{\ell=0}$ at which $g$'s diverge as $\ell\rightarrow\infty$ is identified with the critical temperature
$t_c$. At $t_c$, the growth of the magnitude of the $g$'s is precisely balanced by the decay in the $\Phi$'s. We do not
extend our calculations below $t_c$. For the parameters corresponding to phases $3^+$, $3^-$ and $9^-$ in
Fig.\ \ref{fig:schematic phase diagram}, the $g$'s remain finite as $\ell\rightarrow\infty$ even if $t_{\ell=0}=0$.

Since $\Phi_0 \neq 0$ for $\delta=0$, a finite $\delta$ is generated under RG even if we set the mixed
couplings $\tilde{g}_\mu=0$ as we assume later on. This is similar to the 1-loop renormalization of the mass (or
transition temperature) in bosonic $\phi^4$ theory \cite {KleinertBook_phi4}, and is ultimately responsible
for the slanted shape of the phase boundaries in Fig.\ \ref{fig:schematic phase diagram} near small $g$.

To extract physical information about the nature of the SB we calculate the ordering susceptibilities,
$\chi^{\mathcal{C}}_i$, for a large number of excitonic, i.e., particle-hole (p-h), and superconducting, i.e.,
particle-particle (p-p) channels $\mathcal{C}$.
The $\mathcal{C}$=p-h considered are $\sum_{\alpha \beta} \int d^2\br \psi_\alpha^\dagger(\br)
O^i_{\alpha \beta} \psi_{\beta}(\br)$, where $O^i = \tau_\mu \sigma_\nu \delta_{\alpha \beta}$ for singlet,
and $O^i = \tau_\mu \sigma_\nu \vec \sigma_{\alpha \beta}$ for triplet channels. The $\mathcal{C}$=p-p
considered are $\tfrac 1 2\sum_{\alpha \beta} \int d^2\br \psi_\alpha(\br) \tilde O^i_{\alpha \beta} \psi_{\beta}(\br)+h.c.$.
The $\chi^{\mathcal{C}}_i$ can be calculated using the methods detailed in Ref.\ \cite{CvetkovicThrockmortonVafek2012} and for TLG it is given by
\be
  \chi_{i}^\mathcal{C}= \frac {|\ms|}{8 \pi} \int_0^\infty  \rmd \ell \sum_{\mu \nu}
    \alpha_{i (\mu \nu)}^\mathcal{C} e^{2 \Omega_i^\mathcal{C}\! (\ell)} \Phi^{(\mu \nu)} (t_\ell, \delta_\ell, r_\ell). \label {chiphpp}
\ee
where
\be
  \Omega^\mathcal{C}_{i} (\ell) = \!\!\sum_{\mu \nu \mu' \nu'}\!\! B_{i, (\mu' \nu')}^{\mathcal{C}, \mu \nu}\!\!
    \int_0^\ell \! \! \rmd \ell' g_{\mu \nu} (\ell') \Phi^{(\mu' \nu')} (t_{\ell'}, \delta_{\ell'}, r_{\ell'}), \label {Omega}
\ee
and constants $\alpha_{i, (\mu \nu)}^\mathcal{C}$ and $B_{i, (\mu' \nu')}^{\mathcal{C}, \mu \nu}$ are given in the
SM. The above equations hold in the absence of the mixed
couplings, i.e., $\tilde g_\mu=0$. This holds at the bare level in any of the model cases we
studied in detail. Moreover, if absent, $\tilde g_\mu$ is not generated by Eq.\ \eqref{eq:gtildeflow}.
In each case our numerical evaluations of the RG flow
equations and the $\chi^{\mathcal{C}}_i$'s benefited from additional analytical study of the flow at asymptotically large $\ell$ near $t_c$.
We were able to enumerate all stable flow trajectories (rays) and determine the instabilities and exponents for each of these.

{\it Model 1}: Fig.\ \ref{fig:phase diagrams}a shows the phase diagram in the
forward scattering limit, i.e., only $g_{00} (\ell=0) \neq 0$.
The system exhibits an excitonic (p-h) instability in the spin-singlet $\tau_3 \sigma_2$ channel. This SB
term belongs to $A_{2g}^+$ IR: it breaks lattice {\it mirror} symmetries and $C'_2$ rotations ($180^\circ$ rotations about the
axes perpendicular to principal 3-fold axes). It does {\em not} break the $3-$fold rotational symmetry about the principal
axes \cite{notNematic}, inversion, or time reversal symmetry. We call this mirror-breaking (MB) phase. The spectrum
inside the MB phase is gapless with DP's rotated in the same direction at each $\pm K$ valley.
Our calculations indicate that MB is very sensitive to the presence of even very small back scattering
terms, $g_{\bK 3}/g_{00}$ and $g_{\bK 0}/g_{00}$ $\gtrsim 4 \times 10^{-3}$, in which case
gapped phases are preferred (See Fig.\ \ref {fig:isotherms}).

{\it Model 2}: The results for the Hubbard model are presented in Fig.\ \ref{fig:phase diagrams}b. In this case
the initial couplings in Eq.\ \eqref{eq:Hint} are $g_{03}=g_{00}\sim U$,
$g_{\bK 0}=g_{\bK 3}=\frac{1}{2}g_{00}$ and all other couplings vanish initially.
Due to the $SO(4)$ symmetry of the Hubbard model \cite {CNYangPRL1989,
SCZhangPRL1990,CNYangSCZhangMPLB1990, CvetkovicThrockmortonVafek2012}, the phase boundary
has the same shape for repulsive and attractive on-site interaction $U$. In the former case, $U>0$, we
find the leading excitonic instability to be $O_i=\bbone_{2}\sigma_3\vec{\sigma}$, i.e., a layer antiferromagnet (AF) \cite{MScherer2012}
whose electronic spectrum is gapped. AF breaks time reversal and inversion symmetry. In the latter case,
$U<0$, the leading instability is found to be equally strong for $O_i=\bbone_{2}\sigma_3\bbone_2$ i.e., a spontaneously layer-polarized (LP) state and
$\tilde{O}_i=\tau_1\bbone_2\sigma_2$ i.e., an $s_{++}$-superconducting state (SC). The LP breaks inversion and the SC the
charge $U(1)$ symmetry. Their spectrum is also gapped. Just as in the case of BLG, a mapping
connecting these flows can be found \cite{CvetkovicThrockmortonVafek2012}.
At the transition into the AF phase, all $g$'s diverge along the AF stable ray (see the SM). Due
to the $SO(4)$ symmetry, the asymptotic flow corresponding to the LP+SC instability can be obtained from the AF stable ray.
This is an unstable fixed ray in the $g-$space and is
a consequence of fine tuning. An infinitesimally small interaction that breaks the $SO(4)$ symmetry results either in LP only  or SC only
instability. Numerically, we see this as a diversion of the flow toward one of two distinct stable fixed rays corresponding to these
phases.

{\it Model 3:} Finally, we consider cases of interaction most likely to capture the situation in the experiments where e-e interaction
is screened due split-off bands and gates: backscattering is finite, but small compared to $g_{00}$. This makes
$g_{\bK 3} = g_{\bK 0}$. In the phase diagram, Fig.\ \ref{fig:phase diagrams}c, we set
$g_{\bK 0}/ g_{00} =0.1$ as a representative value. Changing this ratio does not induce
qualitative changes as we show later in Fig.\ \ref{fig:isotherms}.
When $g_{00}>0$ the leading instability is towards AF, the flow following the same stable ray as in the case of the Hubbard model. For
$g_{00}<0$, the leading instability is towards $s_{++}$-SC.
The phase diagram in the case of attractive back scattering (here represented by $g_{\bK 0}/ g_{00} =-0.1$)
is shown in Fig.\ \ref {fig:phase diagrams}d. For the major part of the transition line, we find that the model undergoes a
transition to a SB phase with $O_i=\tau_{3}\sigma_3\vec \sigma$, i.e., (gapped in bulk) quantum spin
Hall phase (QSH) \cite {MScherer2012}.
Only in a small portion of the phase boundary ($g_{00} \lesssim 0.04$) facing the $3^+$ phase  the leading instability
changes to $O_i = \tau_{1} \sigma_3 \bbone_2$ or $O_i = \tau_{2} \sigma_3 \bbone_2$,
the two order parameters being equivalent by  lattice symmetry. This instability leads to the layer polarization density wave (LPDW),
a state where both top and bottom layers exhibit a charge density wave, but the local density of charge has opposite sign on the two layers.
The spectrum of the LPDW phase is gapless.

It is easy to understand why spontaneous rotational SB phases are unfavorable in TLG.
Because the low energy degrees of freedom in TLG reside on the sites directly above/below each
other in the two outer layers, the $120^\circ$ rotations have no effect on the valley-layer matrices
$\tau_{\mu}\sigma_{\nu}$. Therefore, any rotational SB order parameter operator must involve at least one
power of $\bk$, making it less relevant under RG, and therefore less likely to condense, than
the $\bk$-independent operators considered here. This is unlike in BLG, where the two sites are
horizontally displaced, some $\tau_{\mu}\sigma_{\nu}$ transform non-trivially under rotations, and such states can be
favored \cite{VafekYangPRB2010,CvetkovicThrockmortonVafek2012}.

In Fig.\ \ref {fig:isotherms} we examine the influence of back scattering on the transition temperature and the leading instability. We choose
to use non-interacting Hamiltonian parameters \cite {FZhangPRB2010} that yield $\Lambda=0.106/a$, $r_0 = 4.19$,
and $\delta_0=-0.071$, as only for $\delta<0$, the TLG phase
diagram may exhibit reentrant SB phase (see Fig.\ \ref {fig:phase diagrams}a). The relative strength of back to forward scattering,
$g_{\bK 0} / g_{00}$, varies from 0 to 0.1, therefore connecting the left and right edges of this plot to Figs.\ \ref {fig:phase diagrams}a and c,
respectively. With no back scattering and at finite $T$, the critical interaction $g_{00}$ at which the MB phase sets in must be fairly large.
A presence of non-zero $g_{\bK 0}$ quickly suppresses this value, implying that the back scattering helps to enhance $T_c$.
The MB instability is suppressed by back scattering, too, with the AF order preferred whenever $g_{\bK 0} / g_{00} \gtrsim 4 \times 10^{-3}$.
This result suggests that the gapped state observed in TLG experiments \cite {WBaoNatPhys} is most probably the AF state (Fig.\ \ref {fig:isotherms} inset).

The authors acknowledge discussions with B.\ Roy in the early stages of this work and thank R.E.\ Throckmorton for
a critical reading of the manuscript.
This work was supported by the NSF CAREER award under Grant No. DMR-0955561,
NSF Cooperative Agreement No. DMR-0654118, and the State of Florida.

\newpage

\onecolumngrid

\begin {center}
  {\large SUPPLEMENTARY MATERIAL}
\end {center}

\section {Symmetries in ABC trilayer graphene and irreducible representation classifications }

The space group of ABC trilayer graphene (TLG)  lattice symmetries is $P\bar 3 m 1$. The lattice is
invariant under translations by any vector $\bR = m_{+1} \hat b_{+1} + m_{-1} \hat b_{-1}$, where $m_{\pm 1}$ are
integers, and $\hat b_{\pm 1} = a \sqrt 3 / 2 \hat x \pm 3 a / 2 \hat y$ are the unit vectors of the lattice. At $\Gamma$-point,
the point group of lattice symmetries is ${\bf D}_{3d}$; at other high symmetry points of relevance to graphene,
$\pm \bK$, the point group of lattice symmetries is ${\bf D}_3$. The character tables for these two point groups are given
in Table I.

\begin {table}[h]
  \begin {tabular}{| c || c | c | c | c | c | c |}
    \hline
    ${\bf D}_{3d}$ & $e$ & $2 C_3$ & $3 C'_2$ & $i$ & $2 S_6$ & $3 \sigma_d$ \\
    \hline
    \hline
    $A_{1g}$ & 1 & 1 & 1 & 1 & 1 & 1 \\ 
    $A_{2g}$ & 1 & 1 & -1 & 1 & 1 & -1 \\
    $E_{g}$ & 2 & -1 & 0 & 2 & -1 & 0 \\ 
    $A_{1u}$ & 1 & 1 & 1 & -1 & -1 & -1 \\ 
    $A_{2u}$ & 1 & 1 & -1 & -1 & -1 & 1 \\
    $E_{u}$ & 2 & -1 & 0 & -2 & 1 & 0 \\
    \hline
  \end {tabular}
  \hspace {1cm}
  \begin {tabular}{| c || c | c | c |}
    \hline
    ${\bf D}_{3}$ & $e$ & $2 C_3$ & $3 C'_2$ \\
    \hline
    \hline
    $A_{1}$ & 1 & 1 & 1 \\ 
    $A_{2}$ & 1 & 1 & -1 \\
    $E$ & 2 & -1 & 0 \\ 
    \hline
  \end {tabular}
  \caption {Character tables for groups ${\bf D}_{3d}$ and ${\bf D}_3$.}
\end {table}

In addition to obeying the lattice symmetries, we want our Hamiltonian, Eq.\ (1) in the main text, to be invariant under the time reversal, and under the
group of global SU(2) spin rotations. The SU(2) spin rotation invariance is implemented by taking a general Dirac bilinear
 $\sum_{\alpha=\uparrow, \downarrow} \psi_\alpha^\dagger M_{\alpha \beta} \psi_\beta$, and noticing that it is a spin-singlet when
$M = \tau_\mu \sigma_\nu \delta_{\alpha \beta}$, while it is a spin-triplet for $M = \tau_\mu \sigma_\nu \vec \sigma_{\alpha \beta}$.
The non-interacting part of the Hamiltonian may contain only spin-singlet Dirac bilinears, while
the interaction part may have either a product of two spin-singlet or a scalar product of two spin-triplet bilinears.

The tensors formed by the remaining two matrices, $\tau_\mu \sigma_\nu$, are classified according to the irreducible representation
(IR) of the space group they transform under:
\be
  A_{1g, \Gamma}^+ : \bbone_4, \bbone \sigma_1; \qquad A_{2g, \Gamma}^+ : \tau_3 \sigma_2; \qquad A_{2g, \Gamma}^-: \tau_3 \sigma_3; \qquad
    A_{1u, \Gamma}^- : \tau_2 \bbone, \tau_3 \sigma_1; \qquad A_{2u, \Gamma}^+ : \bbone \sigma_3; \qquad
    A_{2u, \Gamma}^-: \bbone \sigma_2; \nonumber \\
  A_{1, \pm \bK}^+ : (\tau_1 \pm i \tau_2) \bbone, (\tau_1 \pm i \tau_2) \sigma_1;
    \qquad A_{2, \pm \bK}^+ :  (\tau_1 \pm i \tau_2) \sigma_3; \qquad A_{2, \pm K}^-:  (\tau_1 \pm i \tau_2) \sigma_2. \qquad \qquad \label {IRs}
\ee
The superscripts in the names of IR's denote whether an $M$ component is even ($+$) or odd ($-$) under time reversal. 

The momentum operators and symmetric tensors they form (up to third order) are classified according to the IR's, too:
\be
  A_{1g, \Gamma}^+ : |\bk|^0 = 1, \quad |\bk|^2 = k_x^2 + k_y^2; \qquad A_{1u, \Gamma}^- : |\bk|^3 \cos 3 \varphi_\bk = k_x (k_x^2 - 3 k_y^2); \qquad
    A_{2u, \Gamma}^- : |\bk|^3 \sin 3 \varphi_\bk = k_y (k_y^2 - 3 k_x^2); \nonumber \\
  E_{g, \Gamma}^+ : |\bk|^2 e^{\pm i 2 \varphi_\bk} = (k_x^2 - k_y^2) \pm i (2 k_x k_y); \qquad
    E_{u, \Gamma}^- : |\bk| e^{\pm i \varphi_\bk} = k_x \pm i k_y, \quad |\bk|^3 e^{\pm i \varphi_\bk} = (k_x^2 + k_y^2) (k_x \pm i k_y). \quad \label {kIRs}
\ee
All Dirac bilinears in TLG belong to one-dimensional IR's. Therefore, they are allowed to apper in a term of
the non-interacting Hamiltonian, Eq.\ (2) in the main text, only if multiplied by the momentum tensor belonging
to the same IR as the bilinear.

\section {$\beta$-function coefficients}

Here we present the expressions for coefficients appearing in Eqs.\ (5-7), (10), and (11) in the main text.

The coefficients appearing in the flow of $\delta$, Eq.\ (5) in the main text, are
\be
  C_{\mu \nu} = 2 \Tr \left ( \left (\sigma_1 \sigma_\nu \right )^2 \right) - 32 \delta_{\mu 0} \delta_{\nu 1}, 
    \qquad \mbox {and} \qquad \tilde C_{\mu} = 8 - 32 \delta_{\mu 0}. \label {tildeCmu} \label {Cmunu}
\ee

Each coefficient $A$ in Eq.\ (7) in the main text has contributions due to random phase approximation (RPA), vertex (V),
and ladder, particle-hole (Lph) and particle-particle (Lpp), types of diagrams
\be
  A_{\mu \nu (\alpha \beta)}^{\kappa \lambda, \kappa' \lambda'} = A_{\mu \nu (\alpha \beta)}^{\kappa \lambda, \kappa' \lambda' ({\rm RPA})} +
    A_{\mu \nu (\alpha \beta)}^{\kappa \lambda, \kappa' \lambda' ({\rm V})} + A_{\mu \nu (\alpha \beta)}^{\kappa \lambda, \kappa' \lambda' ({\rm Lph})} +
    A_{\mu \nu (\alpha \beta)}^{\kappa \lambda, \kappa' \lambda' ({\rm Lpp})}. \label {decomposeA}
\ee
The same kind of decomposition holds for all $\bar A$'s and $\tilde A$'s, appearing in Eqs.\ (6) and (7) in the main text respectively.

The coefficients $A$ are therefore the sum of the following
\be
  A_{\mu \nu (\alpha \beta)}^{\kappa \lambda, \kappa' \lambda' ({\rm RPA})} &=& 2 \delta_{\mu \kappa} \delta_{\nu \lambda}
    \delta_{\mu \kappa'} \delta_{\nu \lambda'} \Tr \left ( \left (  \tau_{\mu} \tau_\alpha \right )^2 \right )
    \Tr \left ( \left ( \sigma_{\nu} \sigma_\beta \right )^2 \right ), \label {A_RPA} \\
  A_{\mu \nu (\alpha \beta)}^{\kappa \lambda, \kappa' \lambda' ({\rm V})} &=& - \frac 1 4 \bigg \lbrack \delta_{\mu \kappa} \delta_{\nu \lambda}
    \Tr \left ( \tau_\kappa \tau_{\kappa'} \tau_\alpha \tau_{\kappa} \tau_\alpha \tau_{\kappa'} \right )
    \Tr \left ( \sigma_\lambda \sigma_{\lambda'} \sigma_\beta \sigma_{\lambda} \sigma_\beta \sigma_{\lambda'} \right ) + \nonumber \\
  && \qquad  \delta_{\mu \kappa'} \delta_{\nu \lambda'}
    \Tr \left ( \tau_{\kappa'} \tau_{\kappa} \tau_\alpha \tau_{\kappa'} \tau_\alpha \tau_{\kappa} \right )
    \Tr \left ( \sigma_{\lambda'} \sigma_{\lambda} \sigma_\beta \sigma_{\lambda'} \sigma_\beta \sigma_{\lambda} \right )
    \bigg \rbrack, \label {A_V} \\
  A_{\mu \nu (\alpha \beta)}^{\kappa \lambda, \kappa' \lambda' ({\rm Lph})} &=& - \frac 1 {16}
    \Tr \left ( \tau_\mu \tau_\kappa \tau_\alpha \tau_{\kappa'} \right ) \Tr \left ( \sigma_\nu \sigma_\lambda \sigma_\beta \sigma_{\lambda'} \right )
    \Tr \left ( \tau_\mu \tau_{\kappa'} \tau_\alpha \tau_{\kappa} \right )
    \Tr \left ( \sigma_\nu \sigma_{\lambda'} \sigma_\beta \sigma_{\lambda} \right ), \label {A_Lph} \\
  A_{\mu \nu (\alpha \beta)}^{\kappa \lambda, \kappa' \lambda' ({\rm Lpp})} &=& - \frac {P_{\alpha \beta}} {32}
    \bigg \lbrack \left ( \Tr \left ( \tau_\mu \tau_\kappa \tau_\alpha \tau_{\kappa'} \right )
    \Tr \left ( \sigma_\nu \sigma_\lambda \sigma_\beta \sigma_{\lambda'} \right ) \right )^2 +
    \left ( \Tr \left ( \tau_\mu \tau_{\kappa'} \tau_\alpha \tau_{\kappa} \right )
    \Tr \left ( \sigma_\nu \sigma_{\lambda'} \sigma_\beta \sigma_{\lambda} \right ) \right )^2 \bigg \rbrack. \label {A_Lpp}
\ee
The prefactor ${P_{\alpha \beta}}$ in the last equation equals $+1$, for $(\alpha \beta) = (01)$, $(30)$, or $(32)$,
and it equals $-1$, for $(\alpha \beta) = (00)$, $(02)$, or $(31)$. One needs to calculate only coefficients with these $(\alpha \beta)$,
since $\Phi^{(\alpha \beta)}$ is non-zero only when $\alpha = 0$ or $3$, {\em and} $\beta = 0$, $1$ or $2$. This is true for all
$A$'s, $\bar A$'s, $\tilde A$'s, $\alpha$'s, and $B$'s, since these coefficients are always summed together with $\Phi^{(\alpha \beta)}$.

The coefficients $\bar A$ are given by
\be
  \bar A_{\mu \nu (\alpha \beta)}^{\kappa, \kappa' ({\rm RPA})} &=& 2 \delta_{\mu \kappa} \delta_{\mu \kappa'}
    \Tr \left ( \left ( \tau_\mu \tau_\alpha \right )^2 \right ) \bigg \lbrack
    \delta_{\nu 0} \Tr \left ( \left ( \sigma_1 \sigma_\beta \right )^2 \right ) + 2 \delta_{\nu 1} \bigg \rbrack, \label {barA_RPA} \\
  \bar A_{\mu \nu (\alpha \beta)}^{\kappa, \kappa' ({\rm V})} &=& - \frac 1 2 \bigg \lbrack \delta_{\mu \kappa}
    \Tr \left ( \tau_\kappa \tau_{\kappa'} \tau_\alpha \tau_\kappa \tau_\alpha \tau_{\kappa'} \right ) + \delta_{\mu \kappa'}
    \Tr \left ( \tau_{\kappa'} \tau_{\kappa} \tau_\alpha \tau_{\kappa'} \tau_\alpha \tau_{\kappa} \right ) \bigg \rbrack
    \bigg \lbrack  \delta_{\nu 0} \Tr \left ( \left (\sigma_1 \sigma_\beta \right )^2 \right ) + 2 \delta_{\nu 1} \bigg \rbrack, \label {barA_V} \\
  \bar A_{\mu \nu (\alpha \beta)}^{\kappa, \kappa' ({\rm Lph})} &=& - \frac 1 8 \Tr \left ( \tau_\mu \tau_\kappa \tau_\alpha \tau_{\kappa'} \right )
    \Tr \left ( \tau_\mu \tau_{\kappa'} \tau_\alpha \tau_{\kappa} \right ) \bigg \lbrack 2 \delta_{\nu \beta}
    \Tr \left ( \left ( \sigma_1 \sigma_\beta \right )^2 \right ) + \left ( \Tr \left ( \sigma_1 \sigma_\nu \sigma_\beta \right ) \right )^2
    \bigg \rbrack, \label {barA_Lph} \\
  \bar A_{\mu \nu (\alpha \beta)}^{\kappa, \kappa' ({\rm Lpp})} &=& - \frac {P_{\alpha \beta}} {16} \bigg \lbrack
    \left ( \Tr \left ( \tau_\mu \tau_\kappa \tau_\alpha \tau_{\kappa'} \right) \right )^2 +
    \left ( \Tr \left ( \tau_\mu \tau_{\kappa'} \tau_\alpha \tau_{\kappa} \right) \right )^2 \bigg \rbrack % \times \nonumber \\
  %&& \qquad \qquad \quad 
    \bigg \lbrack 2 \delta_{\nu \beta} \Tr \left ( \left ( \sigma_1 \sigma_\beta \right )^2 \right ) +
    \Tr \left ( \sigma_1 \sigma_\nu \sigma_\beta \right ) \Tr \left ( \sigma_1 \sigma_\beta \sigma_\nu \right ) \bigg \rbrack. \label {barA_Lpp}
\ee

Finally, the coefficients $\tilde A$ are the sum of the following terms
\be
  \tilde A_{\mu (\alpha \beta)}^{\kappa \lambda, \kappa' ({\rm RPA})} &=& 2 \delta_{\mu \kappa} \delta_{\mu \kappa'}
    \Tr \left ( \left ( \tau_\mu \tau_\alpha \right)^2 \right ) \bigg \lbrack 2 \delta_{\lambda 0} + \delta_{\lambda 1}
    \Tr \left ( \left ( \sigma_1 \sigma_\beta \right)^2 \right ) \bigg \rbrack, \label {tildeA_RPA} \\
  \tilde A_{\mu (\alpha \beta)}^{\kappa \lambda, \kappa' ({\rm V})} &=& - \frac 1 4 \Bigg \lbrace 2 \delta_{\mu \kappa}
    \Tr \left ( \tau_\kappa \tau_{\kappa'} \tau_\alpha \tau_{\kappa} \tau_\alpha \tau_{\kappa'} \right )
    \bigg \lbrack 2 \delta_{\lambda 0} + \delta_{\lambda_1} \Tr \left ( \left ( \sigma_1 \sigma_\beta \right )^2 \right ) \bigg \rbrack + \nonumber \\
  && \qquad \delta_{\mu \kappa'} \Tr \left ( \tau_{\kappa'} \tau_{\kappa} \tau_\alpha \tau_{\kappa'} \tau_\alpha \tau_{\kappa} \right )
    \bigg \lbrack 2 + \Tr \left ( \sigma_1 \sigma_\lambda \sigma_\beta \sigma_1 \sigma_\beta \sigma_\lambda \right ) \bigg \rbrack \Bigg \rbrace, \label {tildeA_V} \\
  \tilde A_{\mu (\alpha \beta)}^{\kappa \lambda, \kappa' ({\rm Lph})} &=& - \frac 1 8 \Tr \left ( \tau_\mu \tau_\kappa \tau_\alpha \tau_{\kappa'} \right )
    \Tr \left ( \tau_\mu \tau_{\kappa'} \tau_\alpha \tau_{\kappa} \right ) \bigg \lbrack 4 \delta_{\lambda \beta} +
    \Tr \left ( \sigma_1 \sigma_ \lambda \sigma_\beta \right ) \Tr \left ( \sigma_1 \sigma_ \beta \sigma_\lambda \right ) \bigg \rbrack, \label {tildeA_Lph} \\
  \tilde A_{\mu (\alpha \beta)}^{\kappa \lambda, \kappa' ({\rm Lpp})} &=& - \frac {P_{\alpha \beta}} {16} \Bigg \lbrace
    \left ( \Tr \left ( \tau_\mu \tau_\kappa \tau_\alpha \tau_{\kappa'} \right ) \right )^2 \bigg \lbrack 4 \delta_{\lambda \beta} +
    \left ( \Tr \left ( \sigma_1 \sigma_\lambda \sigma_\beta \right ) \right )^2 \bigg \rbrack + \nonumber \\
  && \qquad \qquad \qquad 
    \left ( \Tr \left ( \tau_\mu \tau_{\kappa'} \tau_\alpha \tau_{\kappa} \right ) \right )^2 \bigg \lbrack 4 \delta_{\lambda \beta} +
    \left ( \Tr \left ( \sigma_1 \sigma_\beta \sigma_\lambda \right ) \right )^2 \bigg \rbrack \Bigg \rbrace. \label {tildeA_Lpp}
\ee

The particle-hole (p-h) instabilities are defined through Dirac bilinears, $O^i = \tau_\kappa \sigma_\lambda \delta_{\alpha \beta}$ for
singlet and $O^i = \tau_\kappa \sigma_\lambda \vec \sigma_{\alpha \beta}$ for triplet channels. In the case of particle-particle (p-p)
instabilities, Dirac bilinears corresponding to singlet channels are $\tilde O^i = \tau_\kappa \sigma_\lambda \sigma^2_{\alpha \beta}$,
while those corresponding to triplet channels are $\tilde O^i = \tau_\kappa \sigma_\lambda \sigma^a_{\alpha \beta}$, with $a=0$, $1$, or $3$.
The coefficients in the susceptibilities corresponding to one of these channels are
\be
  \alpha_{i, (\mu \nu)}^{{\rm (p-h)}} &=& 2 \Tr \left ( \left ( \tau_\kappa \tau_\mu \right )^2 \right )
    \Tr \left ( \left ( \sigma_\lambda \sigma_\nu \right )^2 \right ), \label {alpha_ph} \\
  \alpha_{i, (\mu \nu)}^{{\rm (p-p)}} &=& - \frac {(-1)^{P_{\mu \nu}}}2 \Tr \left ( \tilde O^i (\tau_\mu \sigma_\nu \bbone)
    \left ( {\tilde O^i - \left ( \tilde O^i \right )^T} \right ) (\tau_\mu \sigma_\nu \bbone)^T \right ) \nonumber \\
  &=& -2 (-1)^{P_{\mu \nu}} \Tr \left ( \left ( \tau_\kappa \tau_\mu \right )^2 \right )
    \Tr \left ( \sigma_\lambda \sigma_\nu \sigma_\lambda \sigma_\nu^T \right ). \label {alpha_pp}
\ee
For the p-h channels, the coefficient $\alpha$ depends only on $\kappa$ and $\lambda$ and not on whether the channel is
singlet or triplet. For the p-p channels, the first line in Eq.\ \eqref {alpha_pp} implies that only channels with antisymmetric $\tilde O^i$'s
can have a non-zero susceptibility. This is already assumed in the second line of the same equation, therefore, it is
applicable for singlet p-p channel susceptibilities only if $\tau_\kappa \sigma_\lambda$ is symmetric, while for
triplet p-p channels it holds only if  $\tau_\kappa \sigma_\lambda$ is antisymmetric. Otherwise, $\alpha^{{\rm (p-p)}} = 0$. Note that
we used $\tau_\mu^T = \tau_\mu$, since $\mu$ can take values $0$ or $3$ only.

Using the same notation, in Eq.\ (11) in the main text we have
\be
  B_{i, (\mu' \nu')}^{{\rm (p-h)}, \mu \nu} &=& 2 \delta_{\mu \kappa} \delta_{\nu \lambda} \delta_i^S
    \Tr \left ( \left ( \tau_\mu \tau_{\mu'} \right )^2 \right ) \Tr \left ( \left ( \sigma_\nu \sigma_{\nu'} \right )^2 \right ) -
    \frac 1 4 \Tr \left ( \tau_\kappa \tau_\mu \tau_{\mu'} \tau_\kappa \tau_{\mu'} \tau_\mu \right )
    \Tr \left ( \sigma_\lambda \sigma_\nu \sigma_{\nu'} \sigma_\lambda \sigma_{\nu'} \sigma_\nu \right ), \label {Bph} \\
  B_{i, (\mu' \nu')}^{{\rm (p-p)}, \mu \nu} &=& - \frac {(-1)^{P_{\mu' \nu'}}} {16} \Tr \left ( \tilde O^i (\tau_\mu \sigma_\nu \bbone)^T
    (\tau_{\mu'} \sigma_{\nu'} \bbone)^T \left ( {\tilde O^i - \left ( \tilde O^i \right )^T} \right ) (\tau_{\mu'} \sigma_{\nu'} \bbone )
    ( \tau_\mu \sigma_\nu \bbone ) \right ) \nonumber \\
  &=& - \frac {(-1)^{P_{\mu' \nu'}}} 4 \Tr \left ( \tau_\kappa \tau_\mu^T \tau_{\mu'} \tau_\kappa \tau_{\mu'} \tau_\mu \right )
    \Tr \left ( \sigma_\lambda \sigma_\nu^T \sigma_{\nu'}^T \sigma_\lambda \sigma_{\nu'} \sigma_\nu \right ). \label {Bpp}
\ee
The symbol $\delta_i^S$ takes value 1 if $O^i$ belongs to a singlet channel and $0$ if it does not. Like before, $B^{\rm (p-p)}$
vanishes if $\tilde O^i$ is a symmetric matrix. The second line in Eq.\ \eqref {Bpp} holds only for antisymmetric p-p channels.

\section {Asymptotic behaviour of the $\Phi$'s and the stable rays}

We can analyse the RG flows exactly in the $\ell \to \infty$ limit. This is a consequence of the fact that $\Phi$ function
in Eqs.\ (8) and (9) in the main text, depend only on three parameters, $\delta_\ell$, $t_\ell$, and $\gamma_\ell$, one of which
will diverge faster than the others. Since $\gamma_\ell$ flows only by its engineering dimension $-1$, Eq.\ (4) in the
main text, the dominant diverging parameter is either $t_\ell$ or $\delta_\ell$. That single parameter determines the asymptotic
behaviour of the $\Phi$'s.

In all the flows that we studied numerically, where $0 \le t_c \lesssim 0.1$, we always found
$\delta_\ell$ to diverge faster than $t_\ell$ at the phase transition. This is the case we present here.
These results are applicable to $T=0$ flows, too, since there is only one diverging parameter, $\delta_\ell$, in that case. We concentrate here on
the case with no mixed interaction terms, $\tilde g_\mu = 0$, at the bare level, so none are generated during the
entire RG flow. 

Under the assumption that $\lim_{\ell \to \infty} t_\ell / \delta_\ell = 0$, we have
\be
  \Phi^{(00)} = \frac {-2}{| \delta |} + O (\delta)^{-2}, \qquad \Phi^{(01)} = \frac {2}{| \delta |} + O (\delta)^{-2}, \label {PhiLargeell}
\ee
while $\Phi^{(\mu \nu)} = O (\delta)^{-2}$ for all other $\Phi^{(\mu \nu)}$'s. Also $\Phi_0 = {\rm sgn} (\delta) + O (\delta)^{-1}$.
These properties allows us to write the flow equations in the large $\ell$ limit, and from there
find all the possible stable rays along which the couplings diverge. 

A ray in the coupling space is defined by the following unit vector,
\be
  z_{\mu \nu} (\ell) = g_{\mu \nu} (\ell) / \sqrt { \sum_{\kappa=0, 3, \bK} \sum_{\lambda=0}^3 g_{\kappa \lambda} (\ell) }, \label {zmunu}
\ee
where the summation in the denominator goes only over 12 independent coupling constants,
thus $g_{\bK \lambda} = g_{1 \lambda}$ or $g_{2 \lambda}$, for each $\lambda$, the two being the same according to the symmetry. 
A stable {\em diverging} ray is given by a unit vector $y_{\mu \nu}$  such that it is a fixed point for the asymptotic flow of $z_{\mu \nu}$:
\be
  \lim_{\ell \to \infty} \left . \frac {\rmd z_{\mu \nu}}{\rmd \ell} \right |_{z \to y} = 0, \label {fixed}
\ee
while at the same time the stability matrix,
\be
  S_{\mu \nu}^{\mu' \nu'} = \lim_{\ell \to \infty}
    \left . \frac {\partial}{\partial z_{\mu' \nu'}} \left ( \frac {\rmd z_{\mu \nu}}{\rmd \ell} \right ) \right |_{z \to y}, \label {stability}
\ee
must have no positive eigenvalues.

For each particular stable ray found, it can be shown that, at the transition point, both $\delta_\ell$ and $g_{\mu \nu}$'s diverge as
$e^{(2+\eta_\delta) \ell}$, where $\eta_\delta$ follows from Eq.\ (5) in the main text,
\be
  \eta_\delta = \lim_{\ell \to \infty} \frac 1 {\delta_\ell}  \sum_{\mu \nu}
    g_{\mu \nu} (\ell) C_{\mu\nu} \Phi_0 \left ( t_{\ell}, \delta_\ell, r_{\ell} \right ). \label {eta_delta}
\ee
One is then required to check if the stable ray obtained is consistent with the initial assumption of $\delta_\ell$
diverging faster than $t_\ell$. If $\eta_\delta <0$ for any given stable ray, such a solution to Eq.\ \eqref {fixed}
is violating our assumption and should be disregarded.

Using the outlined procedure, we find twelve solutions, the coupling constant stable rays, $y_{\mu \nu}$,
and present them in in Table II.
\begin {table}[h]
  \begin {tabular}{| r || c | c | c | c | c | c | c | c | c | c | c | c |}
    \hline
    No.\ & $y_{00}$ & $y_{01}$ & $y_{02}$ & $y_{03}$ & $y_{30}$ & $y_{31}$ & $y_{32}$ & $y_{33}$ &
      $y_{\bK 0}$ & $y_{\bK 1}$ & $y_{\bK 2}$ & $y_{\bK 3}$ \\
    \hline
    \hline
    1 & $0$ & $- \frac 1{\sqrt {31}}$ & $0$ & $0$ & $0$ & $0$ &
      {\tiny $-\sqrt {\frac {5 ( 6+ \sqrt {35} )}{62}}$} & {\tiny $\sqrt{\frac {5 ( 6- \sqrt {35} )}{62}}$ } & $0$ & $0$ & $0$ & $0$ \\
    \hline
    2 & $0$ & $- \frac 1{\sqrt {31}}$ & $0$ & $0$ & $0$ & $0$ &
      {\tiny $\sqrt{\frac {5 ( 6- \sqrt {35} )}{62}}$ } & {\tiny $-\sqrt {\frac {5 ( 6+ \sqrt {35} )}{62}}$ } & $0$ & $0$ & $0$ & $0$ \\
    \hline
    3 & $0$ & $- \frac 1{\sqrt {31}}$ & {\tiny  $-\sqrt {\frac {5 ( 6+ \sqrt {35} )}{62}}$ } &
      {\tiny $\sqrt{\frac {5 ( 6- \sqrt {35} )}{62}}$ } & $0$ & $0$ & $0$ & $0$ & $0$ & $0$ & $0$ & $0$ \\
    \hline
    4 & $0$ & $- \frac 1{\sqrt {31}}$ & {\tiny $\sqrt{\frac {5 ( 6- \sqrt {35} )}{62}}$ } &
       {\tiny $-\sqrt {\frac {5 ( 6+ \sqrt {35} )}{62}}$ } & $0$ & $0$ & $0$ & $0$ & $0$ & $0$ & $0$ & $0$ \\
    \hline
    5 & $0.045$ & $-0.3723$ &$0.1877$ & $-0.2825$ & $0.1394$ & $-0.2341$ & $0.6095$ & $-0.3308$ & $-0.2341$ & $0.1394$ & $-0.2825$ & $0.1877$ \\
    \hline
    6 & $0.045$ & $-0.3723$ & $-0.2825$ & $0.1877$ & $0.1394$ & $-0.2341$ & $-0.3308$ & $0.6095$ & $-0.2341$ & $0.1394$ & $0.1877$ & $-0.2825$ \\
    \hline
    7 & $0.045$ & $-0.3723$ & $0.6095$ & $-0.3308$ & $0.1394$ & $-0.2341$ & $0.1877$ & $-0.2825$ & $0.1394$ & $-0.2341$ & $0.1877$ & $-0.2825$ \\
    \hline
    8 & $0.045$ & $-0.3723$ & $-0.3308$ & $0.6095$ & $0.1394$ & $-0.2341$ & $-0.2825$ & $0.1877$ & $0.1394$ & $-0.2341$ & $-0.2825$ & $0.1877$ \\
    \hline
    9 & $0$ & $-\frac{1}{\sqrt{14}}$ & $0$ & $0$ & $\frac{1}{\sqrt{14}}$ & $0$ & $0$ & $0$ & $0$ & $0$ &
      $-\frac{3+\sqrt{15}}{2 \sqrt{14}}$ & $\frac{-3+\sqrt{15}}{2 \sqrt{14}}$ \\
    \hline
    10 & $0$ & $-\frac{1}{\sqrt{14}}$ & $0$ & $0$ & $\frac{1}{\sqrt{14}}$ & $0$ & $0$ & $0$ & $0$ & $0$ &
      $\frac{-3+\sqrt{15}}{2 \sqrt{14}}$ & $-\frac{3+\sqrt{15}}{2 \sqrt{14}}$ \\
    \hline
    11 & $-\frac{1}{2 \sqrt{38}}$ & $-\frac{5}{2 \sqrt{38}}$ & $\frac{\sqrt{15}}{2 \sqrt {38}}$ & $-\frac{\sqrt{15}}{2 \sqrt {38}}$ &
      $\frac{3}{2 \sqrt{38}}$ & $\frac{3}{2 \sqrt{38}}$ & $-\frac{\sqrt{15}}{2 \sqrt {38}}$ & $\frac{\sqrt{15}}{2 \sqrt {38}}$ &
      $-\frac{3}{2 \sqrt{38}}$  & $-\frac{3}{2 \sqrt{38}}$ & $\frac{\sqrt{15}}{2 \sqrt {38}}$ & $-\frac{\sqrt{15}}{2 \sqrt {38}}$ \\
    \hline
    12 & $-\frac{1}{2 \sqrt{38}}$ & $-\frac{5}{2 \sqrt{38}}$ & $-\frac{\sqrt{15}}{2 \sqrt {38}}$ & $\frac{\sqrt{15}}{2 \sqrt {38}}$ &
      $\frac{3}{2 \sqrt{38}}$ & $\frac{3}{2 \sqrt{38}}$ & $\frac{\sqrt{15}}{2 \sqrt {38}}$ & $-\frac{\sqrt{15}}{2 \sqrt {38}}$ &
      $-\frac{3}{2 \sqrt{38}}$  & $-\frac{3}{2 \sqrt{38}}$ & $-\frac{\sqrt{15}}{2 \sqrt {38}}$ & $\frac{\sqrt{15}}{2 \sqrt {38}}$ \\
    \hline
  \end {tabular}
  \caption {Twelve stable rays.}
\end {table}

In Table II, $y_{\mu \nu}$ for rays Nos.\ 5-8 are given by their numerical value, but can be obtained in a
closed form in terms of zeros of a fourth order polynomial. For example, along the stable ray No.\ 8,
the ratio $\omega = y_{02} / y_{03}$ is a particular zero ($\approx -0.5427$) of polynomial
\be
  0 = 7127 \omega^8 + 31470 \omega^7 + 61602 \omega^6 + 91050 \omega^5 + 109022 \omega^4 + 91050 \omega^3 + 61602 \omega^2
    + 31470 \omega + 7127. \label {AFexact}
\ee
This polynomial is symmetric and can be reduced to a fourth order polynomial by a division by $\omega^4$ and
substitution $z = \omega + 1 / \omega$. Therefore,
there is a closed form solution for $\omega$ which is an algebraic number. The other ratios are then related to $\omega$ as
\be
  \frac {y_{00}}{y_{03}} = \frac {9+ 14 \omega +9 \omega^2 - \Xi }{8 (1+\omega )}, \quad
  \frac {y_{01}}{y_{03}} = \frac{-3+22 \omega -3 \omega^2+3 \Xi}{16 (1+\omega )}, \quad
  \frac {y_{30}}{y_{03}} = \frac {y_{\bK 0}}{y_{03}} = \frac {1 + \omega}2, \qquad \nonumber \\
  \frac {y_{31}}{y_{03}} = \frac {y_{\bK 1}}{y_{03}} = \frac{(1-\omega )^2 - \Xi}{8 (1+\omega )}, \quad
  \frac {y_{32}}{y_{03}} = \frac {y_{\bK 2}}{y_{03}} = \frac{(1+3 \omega)^2- \Xi}{16 (1+\omega )}, \quad
  \frac {y_{33}}{y_{03}} = \frac {y_{\bK 3}}{y_{03}} = \frac{(3 +\omega )^2- \Xi}{16 (1+\omega )},
\ee
where $\Xi = ( 33+60 \omega +70 \omega^2+60 \omega^3+33 \omega^4 )^{1/2}$.
In phases Nos.\ 5-7, the values for the ratios are given by a permutation of the No.\ 8 stable ray ratios.

The properties of the instabilities associated with each stable ray are listed in Table III.
We give the name to each phase, list the order parameter of the leading instability, the IR that the order parameter
transforms according to, whether it is even ($+$) or odd ($-$) under the time reversal symmetry ($\Theta$), if it is gapped or not, and
give $\eta_\delta$.
The instabilities with order parameter transforming according to an IR at $\Gamma$ point have a spatially uniform order parameter. The two instabilities with
the order parameter transforming according to IR's at $\pm \bK$ represent {\em lattice commensurate} density waves.
\begin {table}[h]
  \begin {tabular}{| r || l | c | c | c | c | c |}
    \hline
    No.\ & Name & $O_i$ & IR & $\Theta$ & Gapped & $\eta_\delta$ \\
    \hline
    \hline
    1 & Mirror breaking (MB) & $\tau_3 \sigma_2 \bbone$ & $A_{2g, \Gamma}^+$-singlet & + & N & $1/3$ \\
    \hline
    2 & Quantum anomalous Hall (QAH) & $\tau_3 \sigma_3 \bbone$ & $A_{2g, \Gamma}^-$-singlet & - & Y (in bulk) & $1/3$ \\
    \hline
    3 & Inversion breaking (IB) & $\bbone \sigma_2 \bbone$ & $A_{2u, \Gamma}^-$-singlet & - & N & $1/3$ \\
    \hline
    4 & Layer polarized (LP) & $\bbone \sigma_3 \bbone$ & $A_{2u, \Gamma}^+$-singlet & + & Y & $1/3$ \\
    \hline
    5 & Triplet mirror breaking (tMB) & $\tau_3 \sigma_2 \vec \sigma$ & $A_{2g, \Gamma}^-$-triplet & - & N & $0.4622$ \\
    \hline
    6 & Quantum spin Hall (QSH) & $\tau_3 \sigma_3 \vec \sigma$ & $A_{2g, \Gamma}^+$-triplet & + & Y (in bulk) & $0.4622$ \\
    \hline
    7 & Triplet inversion breaking (tIB) & $\bbone \sigma_2 \vec \sigma$ & $A_{2u, \Gamma}^+$-triplet & + & N & $0.4622$ \\
    \hline
    8 & Layer anti-ferromagnet (AF) & $\bbone \sigma_3 \vec \sigma$ & $A_{2u, \Gamma}^-$-triplet & - & Y & $0.4622$ \\
    \hline
    9 & Interlayer current density wave (ICDW) & $\tau_1 \sigma_2 \bbone$ or $\tau_2 \sigma_2 \bbone$ & $A_{2, \pm \bK}^-$-singlet & - & N & $7/16$ \\
    \hline
    10 & Layer polarization density wave (LPDW) & $\tau_1 \sigma_3 \bbone$ or $\tau_2 \sigma_3 \bbone$ & $A_{2, \pm \bK}^+$-singlet & + & N & $7/16$ \\
    \hline
    11 & $s_{++}$-superconducting ($s_{++}$SC) & $\tau_1 \bbone \sigma_2$ (p-p) & $A_{1g, \Gamma}^+$-singlet & + & Y & $7/16$ \\
    \hline
    12 & Interlayer pairing superconducting ($i$SC) & $\tau_1 \sigma_1 \sigma_2$ (p-p) & $A_{1g, \Gamma}^+$-singlet & + & N & $7/16$ \\
    \hline
  \end {tabular}
  \caption {The list of all the possible phases in TLG with their properties.}
\end {table}

For the cases of interaction considered in the main text we found flows to go into one of the following phases:
MB, AF, QSH, LPDW, and $s_{++}$-SC . In the case of the attractive Hubbard interaction, we found the flow to go
toward a ray obtained from the AF stable ray, Table II, phase No.\ 8, by keeping $y_{01}$, $y_{30}$,
$y_{32}$, $y_{33}$, and $y_{\bK 1}$ the same and reversing the signs of all other $y_{\mu \nu}$'s.
This ray is fixed (Eq.\ \eqref {fixed} holds), but unstable and is a consequence of fine tuning of the interaction.


\begin{thebibliography}{99}
\bibitem{ph} In the absence of particle hole symmetry $9^-$ would correspond to a compensated semimental with particle and hole Fermi surfaces.
\bibitem{ehrenfest} The order of the transition refers to the Ehrenfest classification, i.e., it is labeled by the lowest derivative of the free
  energy with respect to the tuning parameter $\Delta$ which is discontinuous at the transition.
\bibitem{notNematic} Therefore the mirror symmetry breaking state should not be confused with Nematic.
\end{thebibliography}
\end{document}